# Snowfall induced by Corona Discharge

Jingjing Ju[1†], Tie-Jun Wang[1†], Ruxin Li[1*], Shengzhe Du[1], Haiyi Sun[1], Yonghong Liu[1,2], Ye Tian[1], Yafeng Bai[1], Yaoxiang Liu[1], Na Chen[1], Jingwei Wang[1], Cheng Wang[1], Jiansheng Liu[1,3], S. L. Chin[4], and Zhizhan Xu[1*]

[1]State Key Laboratory of High Field Laser Physics, Shanghai Institute of Optics and fine Mechanics (SIOM), Chinese Academy of Sciences, No. 390, Qinghe Road, Jiading District, Shanghai 201800, China.

[2]MOE Key Laboratory of Advanced Micro-structured Material, Institute of Precision Optical Engineering, School of Physics Science and Engineering, Tongji University, Shanghai 200092, China.

[3]IFSA Collaborative Innovation Center, Shanghai Jiao Tong University, Shanghai 200240, China.

[4]Center for Optics, Photonics and Laser (COPL), Laval University, Quebec City, QC G1V 0A6, Canada.

*Correspondence to: ruxinli@mail.shcnc.ac.cn, zzxu@mail.shcnc.ac.cn

†These authors contributed equally to this work.

**Abstract:** We demonstrated for the first time the condensation and precipitation (or snowfall) induced by a corona discharge inside a cloud chamber. Ionic wind was found to have played a more significant role than ions as extra Cloud Condensation Nuclei (CCN). 2.25 g of net snow enhancement was measured after applying a 30 kV corona discharge for 25 min. In comparison with another newly emerging femtosecond laser filamentation method, the snow precipitation induced by the corona discharge has about 4 orders of magnitude higher wall-plug efficiency under similar conditions.

PACS numbers: 52.80.Hc, 92.60.jf, 92.60.Nv

**Introduction**

Drought and desertification become a threat to human survival and development. Artificial rainmaking could help by improving the utilization of water resources up in the sky and bring about impressive socioeconomic benefits to societies [1]. The ionization method was first proposed in 1890 by Nikola Tesla for weather modification. Since then, most researchers aimed at using high voltage induced ionization/discharge to create a lot of supplementary cloud condensation nuclei (CCN) to induce precipitation [2-5]. However, mostly condensation was observed but with no precipitation. There were a few reports claiming that the high density plasma (as CCN) generated by corona discharge helped to create dozens of rain showers [6-9]. But meteorologists from the WMO (World Meteorological Organization) disagreed and concluded: "Weather modification technologies that claim to achieve such large-scale or dramatic effects do not have sound scientific basis." [9].

In this work, we report the first experimental observation of corona discharge induced artificial snowmaking in a cloud chamber principally due to the corona induced ionic wind rather than by the ions as extra CCN. Precipitation occurred in relatively large quantity. The ionic wind is essentially due to the acceleration of positive charges along the electric field lines originating from the pointed electrode towards the surface of the grounded plate. Collisions of these ions with air molecules would result in flows of air packets/parcels along the directions of the field. This would mix up the moist air with a large temperature gradient, resulting in a highly super-saturated state which was crucial to the observed snow precipitation. The findings could help clarify eventually controversies on the ionization based weather modulation methods raised before and implement more significant artificial rainmaking or snowmaking in the atmosphere in the future.

**Experimental setup and results**

The experiments were conducted inside a diffusion cloud chamber ($0.5 \times 0.5 \times 0.2$ m$^3$) with the experimental setup schematically shown in Fig. 1a. The cloud chamber was filled with ambient air with the bottom metal plate cooled down to ~-46 $^o$C and a glass cover at the top with room temperature ~20 $^o$C. Pure water was added into a rectangular water tank mounted about 17.5 cm above the cold plate inside the chamber. The chamber was sealed well and surrounded by thermal insulation foam to keep a stable temperature and humidity distribution inside. A copper cylindrical electrode (1.10 cm (bottom diameter) $\times$ 10 cm (length)) with a pointed end (a conic tip (1.1 cm (bottom diameter) $\times$ 1.6 cm (length) with diameter of the tip ~0.5 mm) was fixed on a plastic base had). The electrode was connected to a high voltage (0-100kV) DC power supply. The height of the electrode from the cold plate could be changed. The chamber was grounded safely by connecting the metal cold plate at the bottom to the ground directly (Fig. 1a, top right). The axis of the electrode was parallel to the bottom base plate. A 2.0 W CW diode pumped semiconductor green laser was used as the probe beam propagating at the height of the electrode with a size of 30 mm (height) $\times$ 5 mm (width). The side Mie scattering was recorded by a digital single-lens reflex camera (Nikon D7000, 4928 $\times$ 3264 pixels) with a macro lens (AF 60 mm/2.8D). Inside the cloud chamber, the relative humidity and the typical vertical temperature distribution in the cloud chamber were measured with a ZDR-F20 Humidity Logger and a regular thermistor thermometer, respectively.

In the present experiment, the total experimental duration was fixed at 50 min. During the first 25 min. the cloud chamber was cooled down from room temperature to a stable state with a temperature gradient from -46 $^o$C (bottom) to 20 $^o$C (top). And in the last 25 min. a corona discharge was triggered with the cooling on all the time. As shown in the supplementary video [10], before the corona discharge was triggered, many background particles (ice and water droplets) floated around slowly with an average velocity ~3.0 cm/s and their maximum size was measured to be ~50 μm (Fig. 2a). Once the corona discharge was triggered, bluish-violet fluorescence emitted from the corona discharge was recorded (Fig. 1b, side view). The fluorescence was due to the ionization of nitrogen molecules and by collisional excitation [11, 12]. Fig. 1b shows that the fluorescence emission became stronger as the voltage was increased. This indicated that more plasma was generated around the tip at higher voltages. The lower part of the fluorescence emission was enhanced more than the upper part and it was bent gradually towards the bottom cold plate as the voltage increased. Meanwhile, a strong plasma channel (a bright fluorescing stem) started to appear along the central axis of the electrode (Fig. 1b). The length of the plasma channel increased with the increase of the applied voltage. Along this plasma channel, a narrow jet-like (electric) wind blows from the electrode forward and towards the cold plate (Fig. 1) [13] as evidenced by the direction of the flow of small packets of fog.

An anti-clockwise vortex appeared within the slightly turbulent air below the electrode at ~2.6 cm from the grounded cold bottom plate when the voltage was increased to ~9.3 kV (Fig. 2A and supplementary video). The vortex was close to the cold plate with a diameter of ~1.5 cm. Packets of fog were formed and together with large size particles they followed the vortex motion below the electrode, with an average velocity of ~7.2 ±2.2 cm/s. Generally, the particles moved away from the electrode and many of them 'dropped' quickly onto the grounded cold plate. This is an indication of the ionic wind following the electric field direction towards the ground. As the voltage increased, both the velocity and size of the vortex increased gradually (Fig. 2 and supplementary video). Packets of fog started to appear around the electrode at voltages higher than 14.5 kV. At 20.5 kV, the velocity of the particles reached up to 30±13 cm/s (Fig. 2b). At the center of the vortex, the air flow ran much faster with a detectable particle velocity of ~75 cm/s (20.5 kV). Large size particles with diameters up to 200-300 μm were captured by the camera (as shown in the elliptical zone in Fig. 2a IV). Above 20.5 kV, mostly turbulent air was created. A maximum velocity of the large size particles dragged by the ionic wind reaching up to about ~1.0 m/s was estimated at 30 kV (with electrode height of 3.0 cm).

The violent air flow motion around the electrode mixed the air up in the region between the electrode and the cold plate where there was a temperature gradient. At the end, instead of a uniform and loose distribution of background snowflakes on the cold plate (when no corona discharge was involved as shown in Fig. 3a), a thick, dense snow pile was observed below the electrode at 10 kV. It had a leaf-shape (Fig. 3b) very similar to the enlarged projection of the corona discharge onto the cold plate. It occupied principally an area of ~7.8 cm × 5.3 cm below the electrode. The snowflakes were more like ice particles and had an average size of ~ 0.9± 0.45 mm (Fig. 3d). This was much smaller than the average size of the background snowflakes (2.3 mm ± 0.26 mm) (Fig. 3c). The size of the snow/ice particles close to the electrode was even smaller than that far away from the electrode.

The voltage on the electrode was varied from zero to 30 kV with the electrode rose a bit from the height of ~2.6 cm to 3.0 cm (in order to avoid breakdown at 30 kV). The total snow weight was measured by collecting all the snow covering the whole cold plate. As shown in Fig. 3E, the snow weight increased from 0.99 ± 0.45 g at zero volt (background) to 3.25± 0.5 g at 30 kV which was 3-4 times the background snow weight. Since both the snow weight (Fig. 3e) and the air flow velocity (Fig. 2b) increased with the applied voltage, one could conclude that the larger the air flow motion was, the more snow was formed.

The height of the electrode relative to the cold plate was also varied with the voltage being fixed at 10.0 kV (Fig. 4). Vortex/turbulence was found always appearing between the electrode and the cold plate. The velocity of the air flow dragged by the ionic wind near the cold plate was observed to be decreasing a little bit with increasing electrode height. When the electrode height was at 2.0 cm, the total snow weight obtained was 1.12 ± 0.3 g. It increased gradually with the increase of electrode height, even the relative humidity around the electrode decreased. When the electrode was raised to a height of 10.0 cm, i. e. in the middle of the chamber (20.0 cm height), the snow weighed about 1.62 ± 0.35 g (Fig. 4a).

**Discussions**

In Fig. 5, a few typical electric field lines issuing from the positive electrode were drawn, with the cathode, the metallic cold plate, being grounded. All electric field lines (such as $E_1$, $E_2$, and $E_3$) started perpendicular to the cone surface of the HV electrode. Due to the sharp curvature of the cone tip, the potential gradient between the positive electrode and distributed cold plate was not uniform. The change of potential U close to the HV electrode was much more rapid than that far away from the electrode, i.e. the potential drop ΔU along the electric filed lines within a certain distance d was bigger near the HV electrode. Therefore, closer to the high voltage (HV) electrode, the electric field E = ΔU/d was much stronger; hence, the density of plasma generated through avalanche ionization was higher and the intensity of fluorescence emitted from recombination of plasma was stronger (as shown in Fig. 1b). Meanwhile, electric field lines (such as $E_1$ and $E_2$) end vertically onto the surface of the conducting cold plate according to the law of electrostatics. As a result, the field distribution around the electrode was not symmetric to the electrode axis, with the lower part being stronger than the upper part (as shown in Fig. 1b, the 20.5 kV case).

High density of plasma was generated once the field was high enough to trigger the avalanche ionization around the sharp metallic tip (anode). Electrons were accelerated along the electric field lines towards the anode. Positive ions would flow along the E-field lines towards the cold plate. This would constitute parcels or clouds of ions or ionic wind. They would blow strongly along the plasma stem from right to left in the view of Fig. 1b and Fig. 2a [13] and stir up an anti-clockwise air vortex. The faster the central stream of the ion motion was, the larger the vortex would be in size. It would eventually degenerate into turbulence as we saw in the experiment at higher voltages.

Note that in the observation (video), most of the air flow in this experiment had a tendency to flow towards the grounded cold plate; this is an indication of the electric field line perpendicular to the ground plate followed by the ionic clouds or ionic wind. The kind of air flow (updraft or two symmetric vortices) observed in the filament case [16, 21-23] due to the thermal effect (convection) was absent or too weak to be observable. We could thus conclude that thermal effect of the plasma generated by the corona discharge might have contributed to the vortex/turbulence motion through convection but not in a dominant way.

The turbulence/vortices dragged by the ionic wind would mix up the region between the cold plate and the electrode. This would, in our experimental condition, create a super-saturation region. Condensation/precipitation through natural CCN (background CCN and background droplets of water and ice/snow) as well as the ions as CCN would follow. Hence, we saw puffs of cloud flowing from the electrode's tip outward and then downward towards the cold plate. Some puffs simply flew from the electrode towards the cold plate not far from the tip because it followed a shorter field line (e.g. $E_2$). Some puffs flew a longer distances because it followed a longer field line (e.g. $E_3$).

The stronger the air flow dragged by the ionic wind was, the more efficient the mixing of moist air with a temperature gradient would be. Therefore, a much more stable super saturated state would be sustained with high collision probability between the CCN's and the water molecules. The final quantity of snow would be large. In the present experiment, this "good" mixing of moist air was supposed to occur mainly in the area with strong ionic wind generated. This area corresponded to the area of corona discharge where strong electric field lines occurred. The leaf-liked shape of the snow pile which looked like the projection of the corona discharge on the cold plate (Fig. 3b) was captured experimentally. It indicated that much of the final precipitation onto the cold plate was from the corona discharge zone (Fig. 3b).

In the area away from the corona discharge where much less electric field lines could reach, the corresponding ionic wind would be too weak to stir up the air. Therefore in the present experiment, outside the zone where ionic wind generated by the corona discharge was strong, snow formation was much less.

The strong field region around the electrode would not change when we increased the electrode height relative to the cold plate with a fixed applied voltage. The total number of positive charges on the electrode mostly concentrated around the tip $+\Delta Q$ and the total negative charges $-\Delta Q$ distributed on the large grounded surface should be the same i. e. $|+\Delta Q| = |-\Delta Q|$. That is to say, the ionic wind around the sharp electrode would be similar for different heights. But increasing the electrode height relative to the cold plate would allow the mixing of a larger (thicker) humid zone below the electrode. For example, when the electrode height was fixed at 2.0 cm, the temperature (T) and relative humidity (RH) around the electrode was measured to be about T = -16 °C and RH = 85.3%. (Fig. 4b point $B_1$) The ionic wind issued from the electrode tip dragged moist air around the electrode towards the grounded cold plate, mixed with the cold moist air there which had a temperature of about T = -46 °C and relative humidity RH = 100% (Fig. 4b point A). Using the same equations given in [14-16] (Fig. 4b), namely:

$$\log(P_{ice}) = \log(611.14) - 9.096853\left(\frac{T_t}{T} - 1\right) - 3.566506 \log\left(\frac{T_t}{T}\right)$$

$$+ 0.876812\left(1 - \frac{T}{T_t}\right) \qquad (180K < T < 273.16K)$$

$$\log(P_{liq}) = \log(611.14) + 10.79574\left(1 - \frac{T_t}{T}\right) - 5.0280 \log\left(\frac{T}{T_t}\right) + 1.50475 \cdot 10^{-4}\left(1 - 10^{-8.2969\left(\frac{T}{T_t} - 1\right)}\right) + 0.42873 \cdot 10^{-3}\left(10^{4.76955\left(1 - \frac{T_t}{T}\right)} - 1\right) \qquad (273.15K < T < 373.15K)$$

where $P_{ice}$ and $P_{liq}$ are saturated vapor pressure relative to ice and liquid water, respectively, $T_t$ refers to the triple point 273.16 K and $T$ is the temperature of moist air in Kelvin unit. The linearly mixed air had an average temperature of -31.0 °C and RH = 217% (Fig. 4b point $C_1$) i.e. super-saturated with a saturation ratio of 2.17. If the height of the electrode was 10.0 cm, the temperature and relative humidity around the electrode tip was T = 3.5 °C, RH = 81% respectively (Fig. 4b point $C_2$). It mixed with the moist air near the cold plate (Fig. 4b point A) and gave a saturation ratio of 3.21 (Fig. 4b point $C_2$). As shown in Fig. 4, the mixed air for different electrode heights was always super-saturated (Fig. 4a green line). This high saturation ratio would naturally enhance the quantity of precipitation as Fig. 4a (red line) shows.

The acidity of the snow induced by the corona discharge was measured roughly by pH test papers; it showed that the snow was almost neutral. To confirm this observation more precisely, the concentration of $NO_3^-$ in the snow was measured by an ion chromatograph (DionexICS-5000+); it was indeed very low, ~ 0.6 ± 0.05 ppm. This was compared with the $NO_3^-$ concentration ~0.2 ± 0.03 ppm in the background. One could conclude that little binary $H_2O$- $HNO_3$ was produced during the corona discharge. The CCN generated by the corona discharge would be mainly from the plasmas, which could charge the neutral molecules (such as $O_2^-$ etc.) and enhanced their ability of picking up water molecules around [17-19]. However, there were already water/ice droplets and aerosols in the micrometer (μm) scale or larger (cloud) floating in the cloud chamber before the corona discharge was triggered. Such pre-existing droplets and aerosols would compete with the corona discharge generated ions, charged molecules such as $O_2^-$, and the little amount of binary chemicals such as $HNO_3$ for the capture of water molecules and grow in size. Meanwhile, those pre-existing CCN and water/ice droplets would also collide with and adsorb the ions and chemicals rendering them even more active. We would thus expect that the pre-existing larger size CCN's and water/ice droplets, including those activated by the corona induced ions and chemicals, would be the principal source for precipitation, i.e. the background particles were more important than the corona discharge induced charged molecules or chemicals [20]. These background CCN and water/ice droplets would grow efficiently and precipitate in the sustained super-saturated environment which was achieved and maintained for a long period of time in the cloud chamber.

From the experimental results and analyses presented above, we could conclude that to create precipitation in the present experiment, the following four conditions should be satisfied so as to create a sustained super-saturation environment: a high humidity environment, existing CCN, large temperature gradient and air mixing. These four conditions were also satisfied in our previous work by firing 1 kHz femtosecond laser pulses into the same cloud chamber [21-23]. Precipitation occurred in the filamentation case [21-23]. However, the weight of snow induced by filamentation was measured to be only about 13.0 mg under the highest available laser power (about 9 W) after 30 min. of irradiation [21]. This was more than 2 orders of magnitude less than the current case of about 2.25 g of net snow enhancement after applying 30 kV for 25 min. The electric power used in this corona discharge was very small, about 1-10 W in the present work (~0.27 mA with 20 kV applied). This power was similar to the optical power used in the femtosecond filamentation method. Moreover, by taking into account the wall-plug efficiency of femtosecond laser pulses which is usually less than 1% from electrical energy to optical energy, the overall wall-plug efficiency of corona discharge induced snow-making would be 4 orders of magnitude higher than that of the femtosecond laser filamentation method under very similar experimental conditions. This dramatic enhancement of efficiency was due to the constant and strong ionic wind created by the corona discharge mixing up air flow with large temperature gradient and sustaining supersaturated state with much larger space scale than the femtosecond filamentation did. This also shows that the prospect of using corona discharge to induce precipitation is excellent whatever the final outcome of the two techniques will be in the long run [11, 24]. Other similar but very different works were reported before as mentioned at the beginning of this paper [2-8]. Based on our current laboratory findings, we believe that if the four conditions for precipitation given above were satisfied simultaneously at that time, they would have succeeded.

Considering outdoor experiments of this corona discharge method, high voltage towers might have a limit in height for reaching a high altitude where clouds have a significant temperature gradient. To solve this problem, we could mount them on high altitude mountains or make use of a recent finding of ours [11] in which a femtosecond laser filament could guide the high voltage from one end of the filament to the other. This is equivalent to an extension of the electrode. Since such filaments could be generated to a long distance in air [25-27], it would be possible to guide high voltage corona discharge to a long distance in air. Extension to outdoor experiments should be feasible. We are now preparing such an experiment.

**Conclusions**

We demonstrated for the first time that a corona discharge could induce condensation and precipitation (snowfall). Precipitation occurred in relatively large quantity by turning on a static corona discharge inside a laboratory diffusion cloud chamber for 25 min. under a potential difference larger than 10 kV between a pointed positively charged electrode and the ground plate. We attribute this phenomenon to the precipitation in a sustained super-saturation environment aided mainly by the ionic wind from the corona discharge.

**Acknowledgements:** We acknowledge the National Basic Research Program of China (2011CB808100), the National Natural Science Foundation of China (11404354, 11425418, 61475167, 61221064, 60921004, 11204329, and 11174305), the State Key Laboratory Program of the Chinese Ministry of Science and Technology, and Chinese Academy of Sciences. J. Liu acknowledges the support of Shanghai Science and Technology Talent Project (Nos. 12XD1405200). T. J. Wang acknowledges the support from 100 Talents Program of Chinese Academy of Sciences and Shanghai Pujiang Program. S. L. Chin acknowledges the support of Laval University, Quebec City, Canada. Fruitful discussions with Prof. X. Guo and Prof. H. Xu from Chinese Academy of Meteorological Sciences are acknowledged sincerely.

**References and Notes:**

1.  J. Qiu, D. Cressey, Nature **453**, 970–974 (2008).

2.  B. Vonnegut, K. Maynard, W. G. Sykes, and C. B. Moore, J. Geophys. Res. **66**, 823-830 (1961).

3. B. Vonnegut, C. R. Smallman, C. K. Harris, and W. G. Sykes, J. Atmos. Terr. Phys. **29**, 781-792 (1967).

4. C. B. Moore, B. Vonnegut, T. D. Rolan, J. W. Cobb, D. N. Holden, R. T. Hignight, S. M. Mcwilliams, and G. W. Cadwell, Science **26**, 1413-1416 (1986).

5. C. B. Moore, B. Vonnegut, E. A. Vrablik, and D. A. McCaig, J. Atmo. Sci. **21**, 646-665, (1964).

6. M. Reznikov, IEEE Trans. Industry Appl. **51**, 1137-1145 (2015).

7. S. Beare, R. Chambers, and S. Peak, Statistical modelling of rainfall enhancement. University of Wollongong, Working paper 14-09, 30p. (2009).

8. D. R. Nevada, Dailymail. **1,** 43-45, (2011).

9. A. J. Davis and B. Handwerk, National Geographic **01**, 300-305 (2011).

10. **Supplementary video:** Air flow motion induced by corona discharge in a cloud chamber, with high voltage applied on the electrode being varied from 0 to 20.5 kV. The electrode was set horizontally at a height of ~2.6 cm relative to the cold plate.

11. F.W. Jr. Peek, Jour. Franklin Inst. **197**, 1-44, (1924).

12. T. J. Wang, Y. Wei, Y. Liu, N. Chen, Y. Liu, J. Ju, H. Sun, C. Wang, H. Lu, J. Liu, S. L. Chin, R. Li and Z. Xu, Sci. Rep. **5**, 18681 (2015).

13. B. Ladenburg, Ann. Physik **4**, 863-897 (1930).

14. D. M. Murphy, T. Koop, Q. J. R. Meteorol. Soc. **131**, 1539 (2005).

15. This is based upon an original idea of T. Leisner from the Karlsruhe Institute of Technology (Germany), in a private discussion with one of us (SLC) in 2013.

16. J. Ju, T. Leisner, H. Sun, A. Sridharan, T. J. Wang, J. Wang, C. Wang, J. Liu, R. Li, Z. Xu and S. L. Chin, Appl. Phys. B **117**, 1001-1007 (2014).

17. R. G. Harrison, Space Sci. Rev. **94**, 381-396 (2000).

18. K. S. Carslaw, R. G. Harrison, and J. Kirkby, Science **298**, 1732-1737 (2002).

19. M. Kulmala, Science **302**, 1000-1001 (2003).

20. U. Dusek, G. P. Frank, L. Hildebrandt, J. Curtius, J. Schneider, S. Walter, D. Chand, F. Drewnick, S. Hings, D. Jung, S. Borrmann, and M. O. Andreae, Science **312**, 1375-1378 (2006).

21. J. Ju, J. Liu, C. Wang, H. Sun, W. Wang, X. Ge, C. Li, S. L. Chin, R. Li, and Z. Xu, Opt. Lett. **37**, 1214-1216 (2012).

22. J. Ju, H. Sun, A. Sridharan, T. J. Wang, C. Wang, J. Liu, R. Li, Z. Xu, and S. L. Chin, Phys. Rev. E **88**, 062803 (2013).

23. J. Ju, J. Liu, H. Liang, Y. Chen, H. Sun, Y. Liu, J. Wang, C. Wang, T. J. Wang, R. Li, Z. Xu, and S. L. Chin, Sci. Rep. **6**, 25417 (2016).

24. P. Rohwetter, J. Kasparian, K. Stelmaszczyk, Z. Hao, S. Henin, N. Lascoux, W. M. Nakaema, Y. Petit, M. Queißer, R. Slamé, E. Salmon, L. Wöste, and J-P Wolf, Nat. Photon. **4**, 451-456 (2010).

25. P. Rairoux, H. Schillinger, S. Niedermeier, M. Rodriguez, F. Ronneberger, R. Sauerbrey, B. Stein, D. Waite, C. Wedekind, H. Wille, L. Wöste, and C. Ziener, Appl. Phys. B **71**, 573-580 (2000).

26. J. Kasparian, M. Rodriguez, G. Méjean, J. Yu, E. Salmon, H. Wille, R. Bourayou, S. Frey, Y.-B. André, A. Mysyrowicz, R. Sauerbrey, J.-P. Wolf and L.Wöste, Science **301**, 61-64 (2003).

27. S. L. Chin, S. A. Hosseini, W. Liu, Q. Luo, F. Théberge, N. Aközbek. A. Becker, V. P. Kandidov, O. G. Kosareva, and H. Schroeder, Can. J. Phys. **6**, 2-63 (2007).

**Figure captions**

**Fig. 1**. (a) A schematic diagram of the experimental setup. (b) Corona discharge under different high voltages. The electrode height was fixed at ~2.6 cm relative to the cold bottom plate.

**Fig. 2.** (a) Side images of corona discharge induced air flow at different high voltage on the electrode. The electrode was set at about 2.6 cm above the grounded cold bottom plate. White dotted lines indicated the moving directions of the air flow. (b) Airflow velocity vs. the voltage applied on the electrode.

**Fig. 3.** (a) Background snow formation on the cold bottom plate; (b) Snow formation on the cold plate when the corona discharge was turned on at 10 kV inside the chamber. (c-d) Close-up shots of typical snowflakes in background (a) and corona discharge-induced snow formation (b), respectively. (e) Weight of snow induced by corona discharge vs. the voltage applied on the electrode (with electrode height fixed of 3.0 cm).

**Fig. 4.** (a) Snow weight vs. different heights of electrode relative to the cold bottom plate (red curve) (the voltage applied was fixed at 10 kV) and the corresponding calculated saturation ratio of mixed air (green curve); (b) Saturated vapor density $\rho_s$ of mixed air at representative heights of 2.0 cm ($C_1$, green lines) and 10.0 cm ($C_2$, blue lines).

**Fig. 5.** Physical picture of electric field lines, which guided the ionic wind, induced by the positive corona discharge. $E_3$ indicates that it will end on some far away positions on the cold plate or some other ground.

**Fig. 1**. (a) A schematic diagram of the experimental setup. (b) Corona discharge under different high voltages. The electrode height was fixed at ~2.6 cm relative to the cold bottom plate.

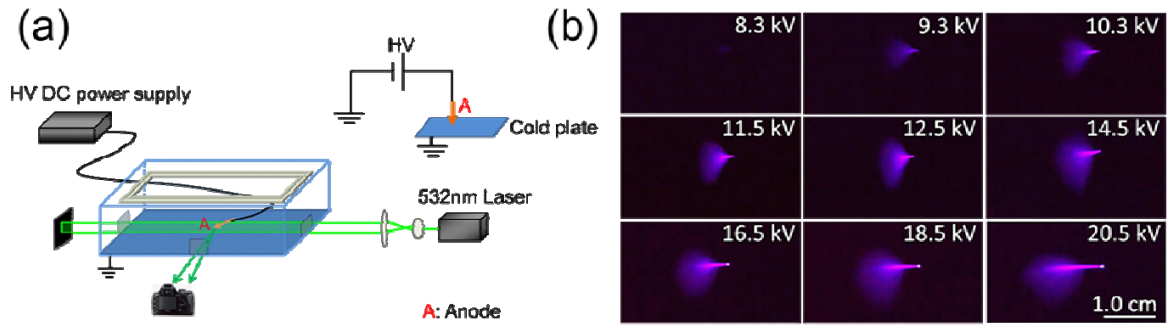

**Fig. 1 by J. Ju *et al.***

**Fig. 2.** (a) Side images of corona discharge induced air flow at different high voltage on the electrode. The electrode was set at about 2.6 cm above the grounded cold bottom plate. White dotted lines indicated the moving directions of the air flow. (b) Airflow velocity vs. the voltage applied on the electrode.

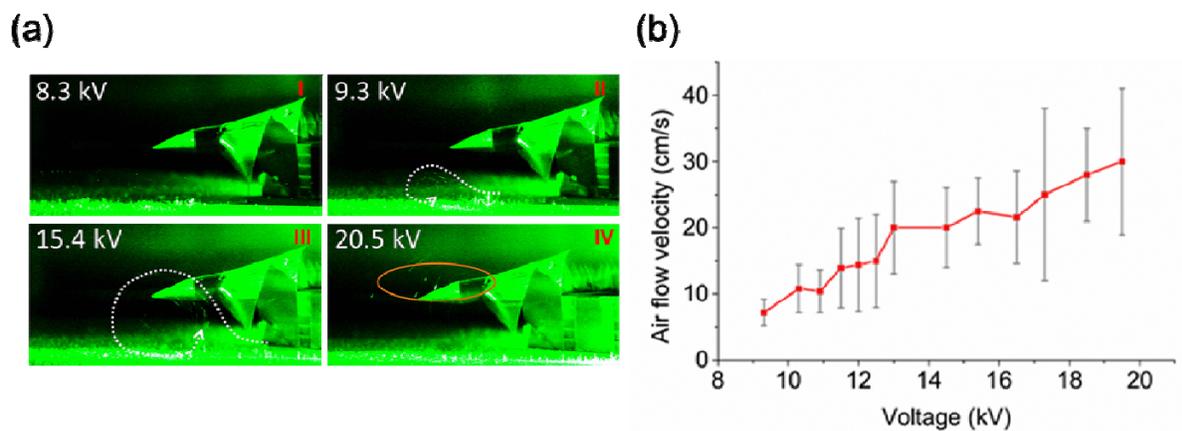

**Fig. 2 by J. Ju *et al.***

**Fig. 3.** (a) Background snow formation on the cold bottom plate; (b) Snow formation on the cold plate when the corona discharge was turned on at 10 kV inside the chamber. (c-d) Close-up shots of typical snowflakes in background (a) and corona discharge-induced snow formation (b), respectively. (e) Weight of snow induced by corona discharge vs. the voltage applied on the electrode (with electrode height fixed of 3.0 cm).

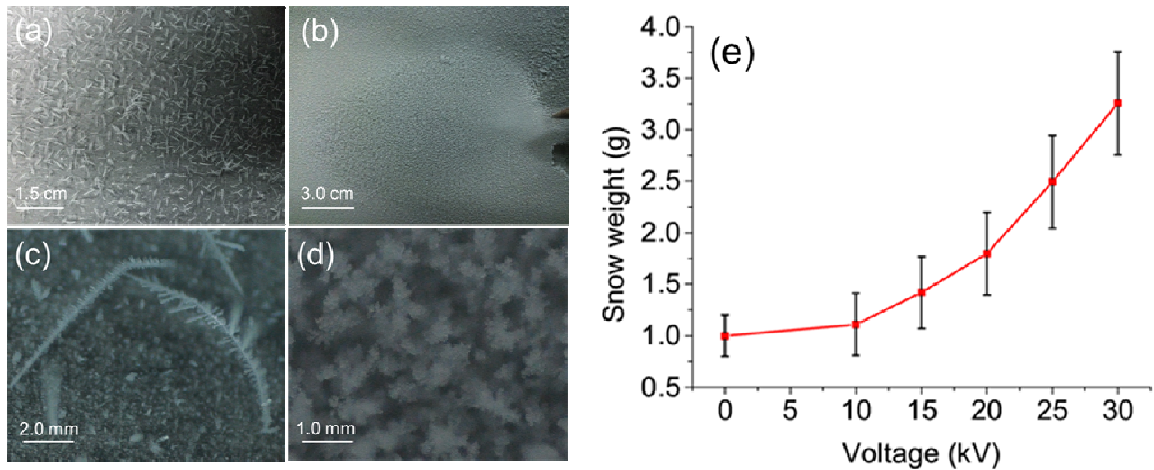

**Fig. 3 by J. Ju *et al.***

**Fig. 4.** (a) Snow weight vs. different heights of electrode relative to the cold bottom plate (red curve) (the voltage applied was fixed at 10 kV) and the corresponding calculated saturation ratio of mixed air (green curve); (b) Saturated vapor density $\rho_s$ of mixed air at representative heights of 2.0 cm ($C_1$, green lines) and 10.0 cm ($C_2$, blue lines).

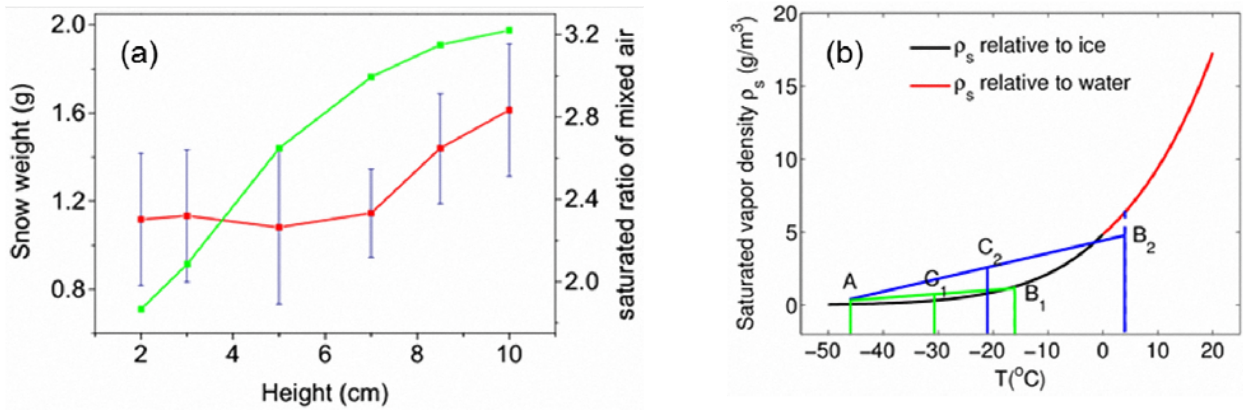

**Fig. 4 by J. Ju *et al.***

**Fig. 5.** Physical picture of electric field lines, which guided the ionic wind, induced by the positive corona discharge. $E_3$ indicates that it will end on some far away positions on the cold plate or some other ground.

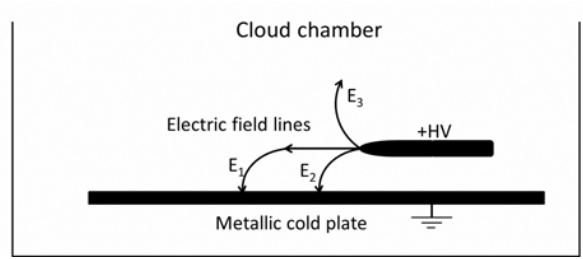

**Fig. 5 by J. Ju *et al.***